%% file: ckm2006_punzi.tex
\documentclass[twocolumn,nofootinbib,amsmath,amssymb,a4paper]{revtex4}

\usepackage{graphicx}
\usepackage{dcolumn}
\usepackage{bm}

\begin{document}
\include{macro}

\newcommand{\BBdpipi}{\ensuremath{\overline{B}^{0}\rightarrow\pi^{+}\pi^{-}}}
\newcommand{\BBdKpi}{\ensuremath{\BBd \rightarrow K^{-}\pi^{+}}}
\newcommand{\Bhhgen}{$B\rightarrow h^{+}h'^{-}$}
\newcommand{\Bhad}{\ensuremath{B}}

\title{\Bhhgen\ modes at CDF}

\author{G. Punzi for the CDF collaboration}
 \email{giovanni.punzi@pi.infn.it}
\affiliation{Universita' di Pisa and I.N.F.N.\\
Largo B. Pontecorvo, 3 - 56127 Pisa, Italy\\}

\begin{abstract}
We review CDF results and prospects on decays of $B$ hadrons in two charged charmless hadrons.
\end{abstract}

\maketitle

\section{Introduction}
The rich production of all types of $B$ hadrons at the Tevatron 
Collider at Fermilab allows studying charmless two--body decays 
both in known and new modes. 
Using a sample of $\int\Lumi dt\simeq 1$~\lumifb\ of data, 
CDF performed a search for all possible modes of neutral bottom 
hadrons in two charged charmless hadrons ($p$, $K$ or $\pi$)~\cite{C-conjugate}. 

The CDFII detector is a multipurpose magnetic spectrometer surrounded by
calorimeters and muon detectors~\cite{CDF}. A silicon micro-strip detector (SVXII) and a cylindrical drift chamber
(COT) immersed in a 1.4~T solenoidal magnetic field
reconstruct charged particles within pseudorapidity
$|\eta | < 1.0$, with a
transverse momentum resolution $\sigma_{p_{T}}/p_{T} \simeq
0.15\%\, p_{T}$/(GeV/$c$). 
This yields a mass resolution of $\simeq 22$ \massmev\ for 
\Bhhgen\ decays, which is an important ingredient of this analysis.
Particle identification information (PID) is obtained from the specific 
energy loss by ionization (\dedx) of charged particles in the COT.
This yields a nearly-constant separation of 1.4 $\sigma$
between pions and kaons of momenta $>$2~GeV/$c$.
A three-level trigger sytem allows selecting events by requiring the 
presence of at least two tracks with large impact parameters relative 
to the beam axis. 

\section{Data sample and reconstruction}

\Bhad\ hadron candidates are initially selected by forming pairs of oppositely-charged 
tracks with $p_{T} > 2$~\pgev , transverse opening-angle 
$20^\circ < \Delta\phi < 135^\circ$, and  $p_{T}(1) + p_{T}(2) > 5.5$~\pgev .
In addition, both tracks are required to have a large transverse impact parameter $d_0$
relative to the $\bar{p}p$ interaction vertex (100 $\mu$m $< d_0 < 1$~mm).
The \Bhad\ candidate is required to point back to
the primary vertex ($d_0(B)< 140$~$\mu$m),  and to have travelled a transverse distance
$L_{xy}(B)>200$~$\mu$m.

Most of the above cuts are implemented at the trigger level. In the 
offline analysis, additional cuts are imposed on isolation $I_{B}$~\cite{isolation} and 
the quality of the fit ($\chi^{2}$) to the 3D decay vertex of the $B$ hadron candidate.

Final selection cuts are determined by an optimization 
procedure, based on minimizing the expected uncertainty of the physics observables to be measured.
Two different sets of cuts are used, optimized respectively for best 
resolution on \acpbdkpi\ (loose cuts), and for best 
sensitivity~\cite{gp0308063} for the discovery of the yet unobserved \BsKpi\ mode (tight cuts).
The looser set of cuts is also used for measuring 
the decay rates of the largest-yields modes, while the 
tighter set is used for the other rare modes.

\begin{figure*}[htb]
\includegraphics[scale=0.35]{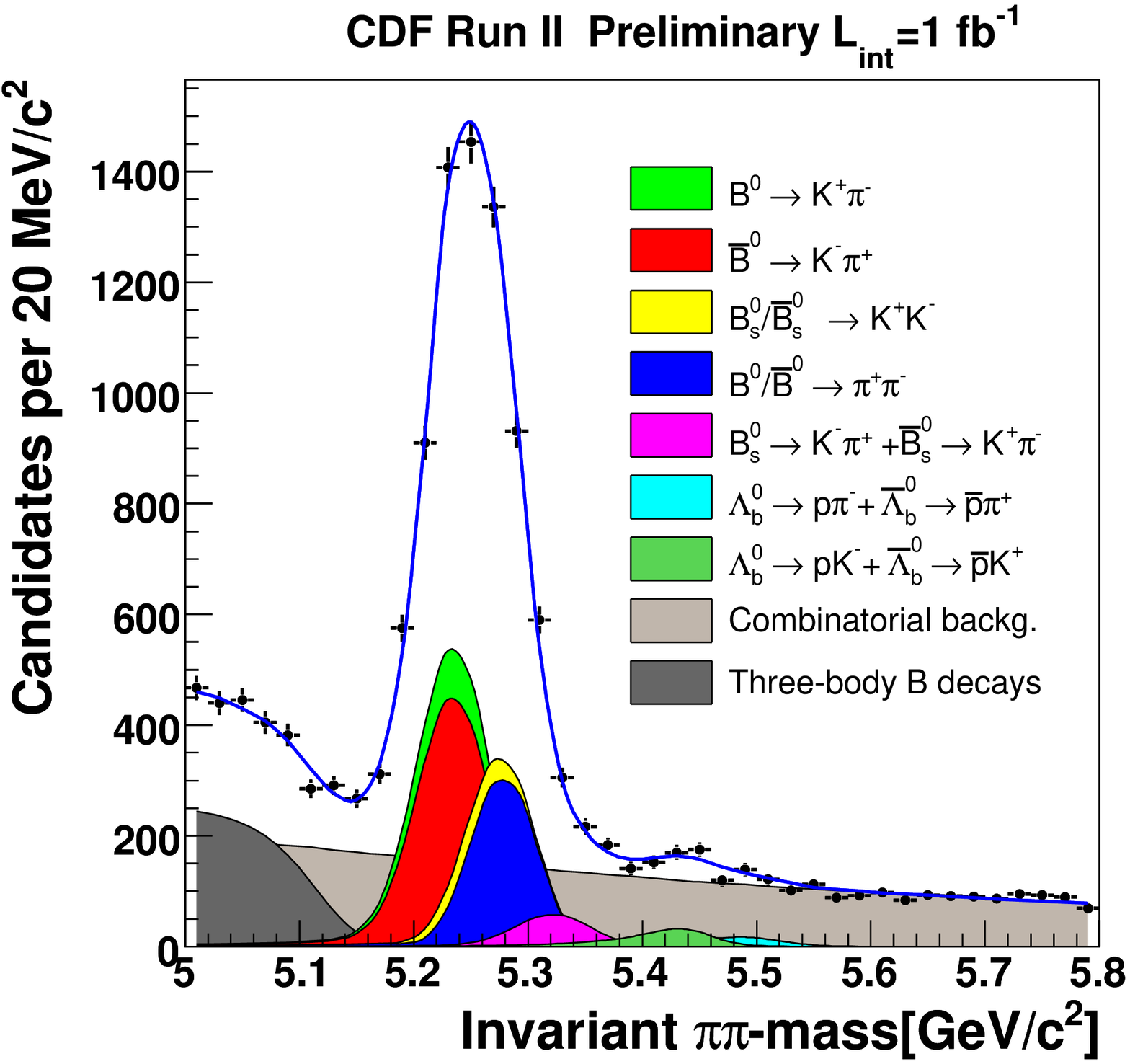}
\includegraphics[scale=0.35]{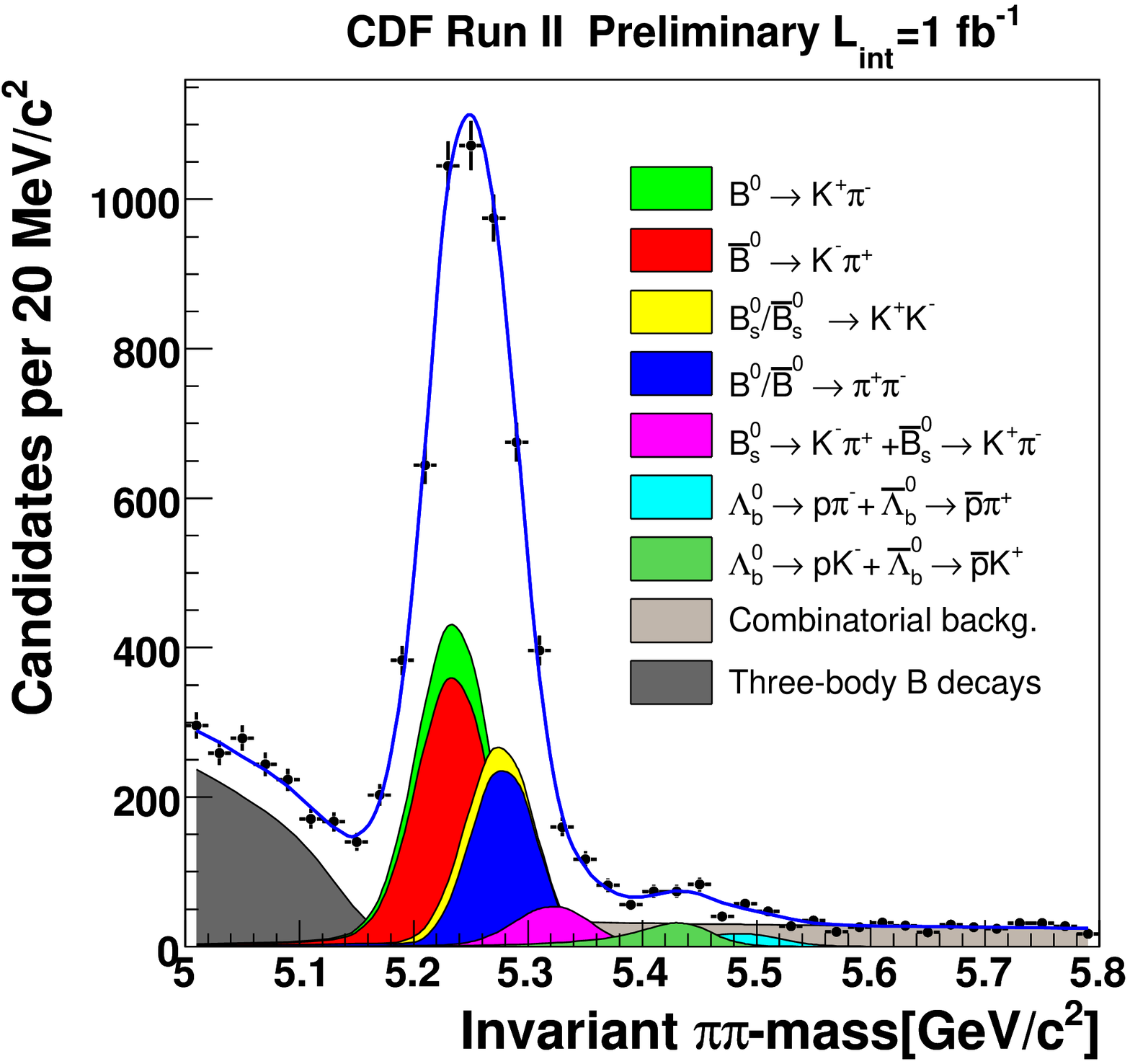}
\caption{Invariant mass distribution
of \bhh\ candidates passing the loose (left) or tight (right) 
selection cuts. The pion mass is assigned to both tracks.
Cumulative projections of the likelihood fit for each mode are
overlaid.}
\label{fig:projections}
\end{figure*}

The invariant mass distribution of the candidates, with an 
arbitrary pion mass assignment to both tracks, shows a single large peak in the $B$ mass range, 
formed by several overlapping modes (\fig{projections}).

\section{Measurement Methodology}
The different modes are statistically separated and individually measured by 
means of an unbinned maximum-Likelihood fit, combining kinematics and PID.
Kinematic information is summarized by three loosely correlated
observables: (a) the mass
$M_{\pi\pi}$ calculated with the pion mass assignment to both
particles; (b) the signed momentum imbalance
$\alpha = (1-p_1/p_2) q_{1}$, where $p_1$ ($p_2$) is the
lower (higher) of the particle momenta, and $q_1$ is the sign of the charge of the
particle of momentum $p_{1}$; (c) the scalar sum of particle momenta $p_{tot}=p_1 + p_2$.
The above variables allow evaluating the mass of the $B$ candidate for 
any mass assigment to the decay products.
PID information is
given by a \dedx\ measurement for each track.
The Likelihood for the $i^{\mathit{th}}$ event is then:
\begin{eqnarray}\label{eq:likelihood}
    \mathcal{L}_i & = & (1-b)\sum_{j} f_j \mathcal{L}^{\mathrm{kin}}_j  \mathcal{L}^{\mathrm{PID}}_j \nonumber \\
                  &   & +  b \left( f_{\rm{A}} \mathcal{L}^{\mathrm{kin}}_{\mathrm{A}}
    \mathcal{L}^{\mathrm{PID}}_{\mathrm{A}}+
   (1-f_{\rm{A}}) \mathcal{L}^{\mathrm{kin}}_{\mathrm{E}}
    \mathcal{L}^{\mathrm{PID}}_{\mathrm{E}}
	\right)
\end{eqnarray}
where index `$\mathrm{A(E)}$' labels the physics (combinatorial) 
background-related quantities, and
index $j$ runs over the twelve possible \Bhhgen\ and \Lbph\ modes and conjugates 
having distinguishable final states (e.g. \BdKpi\ and \BBdKpi\ 
are distinct, while \Bdpipi\ and \BBdpipi\ are treated as a single 
component). The $f_j$ are the signal fractions to be determined by 
the fit, together with the background fraction parameters $b$ and $f_{A}$.

The shape of the mass distribution of each single channel accounts 
for non-Gaussian tails, both from resolution and from emission 
of photons in the final state, which is simulated on the basis of 
analytical QED calculations~\cite{Baracchini-Isidori}. 
The quality of this model was checked on a large 
sample ($\simeq$500k) of \DKpi\ decays.
The mass distribution of the combinatorial background is fit to a smooth 
function, while the physics background is parameterized by an 
`Argus function'~\cite{argus} smeared with our mass resolution. 
Kinematical distributions for the signal are represented by 
analytical expressions, while for the combinatorial background are 
parameterized from the mass sidebands of data~\cite{Beauty06-MM}.

The \dedx\ response of the detector to kaons and pions was calibrated 
from a sample of 1.5M $D^{*+}\to D^0\pi^+
 \to [K^-\pi^+]\pi^+$ decays, where the $D^0$ decay
products are identified by the charge of the $D^{*+}$ pion.
The PID term for the background allows
for pion, kaon, proton, and electron components, which are free to
vary independently in the fit. Background muons are indistinguishable from pions with the
available \dedx\ resolution and are therefore included as a part 
of the pion component.

To avoid the large uncertainties associated to production cross sections 
and reconstruction efficiency, branching fractions are measured 
relative to the \BdKpi\ mode, and then normalized to the world-average value of
\BR(\BdKpi)~\cite{HFAG06}. Upper limits~\cite{F-C} are quoted for modes 
in which no significant signal is observed.

To convert the raw signal fractions returned by the fit into relative branching fractions,
corrections are applied for different efficiencies of trigger and offline selection requirements
for different decay modes. Corrections related to decay kinematics 
are determined from the
detector simulation, while others are measured on data using control samples.
The dominant contributions to the systematic uncertainty come from:
statistical uncertainty on isolation efficiency ratio (for \Bs\ 
modes); uncertainty on the \dedx\ calibration and parameterization; 
and uncertainty on the combinatorial background model.
Smaller systematic uncertainties are assigned for: 
trigger efficiencies; physics background shape and kinematics; $B$ 
meson masses and lifetimes. 

\section{Results}

\subsection{Rare Modes}

The search for 
rare modes is performed using the `tight' selection.
The fit allows for the presence of any component of the 
form \Bhhgen\ or \Lbph\ where $h,h' = K$ or $\pi$, with the yield as a free parameter. 
Final results are reported in Table~\ref{tab:summary}, where $f_{d}$ and $f_{s}$ indicate
the production fractions respectively of \Bd\ and \Bs\
from fragmentation of a $b$ quark in $\bar{p}p$ collisions.

The results provide the first observation of the \BsKpi\ mode, with a 
significance of $8.2 \sigma$, which includes systematic uncertainties 
and is evaluated from Monte Carlo samples of background without signal.
The branching fraction of this mode is significantly sensitive to the value of 
angle $\gamma$; however, predictions obtained 
from different methods differ. Our measurement $\BR(\BsKpi)=(5.0 
\pm 0.75 \pm 1.0)\times 10^{-6}$ is in agreement with the prediction in~\cite{zupan},
but is lower than most other predictions~\cite{B-N,Yu-Li-Cai,Bs-QCDF}.

No evidence is found for modes
\Bspipi\ or \BdKK , in agreement with expectations of 
significantly smaller branching fractions.
The measurement $\BR(\BdKK) = (0.39 \pm 0.16 \pm 0.12)\times
10^{-6}$ has the same precision of the other 
current measurements from \yqs\ experiments~\cite{HFAG06}, although the upper limit is weaker due to 
the observed positive central value.
The \Bspipi\ upper limit is improved with respect to the previous best 
limit, obtained from the analysis of a subsample of the present data~\cite{paper_bhh}.
The sensitivity to both \BdKK\ and \Bspipi\ is now 
close to the upper end of theoretically expected 
range~\cite{B-N,Bspipi-QCDF,Bspipi-PQCD,Bs-QCDF}, 
and it will be interesting to see the results from the larger samples being accumulated.
These modes proceed only through annihilation and exchange diagrams, which
are currently poorly known and a source of significant uncertainty in
many theoretical calculations. Measuring of constraining both these channels is particularly useful
since the physics parameters governing the strength of 
penguin-annihilation can be extracted from their ratio~\cite{Burasetal}.

In the same sample, we also get to observe charmless decays of 
a $B$ baryon for the first time: \Lbppi\ ($6 \sigma$) and \LbpK\ ($11.5 \sigma$).
We measure the ratio of branching fractions of these modes as
$\LbppisuLbpK =  0.66 \pm 0.14 \pm 0.08$, in good agreement with the 
expected range $[0.60,0.62]$ from~\cite{Mohanta-BR}.
Work is in progress towards a measurement of individual branching 
fractions and \CP\ asymmmetries, that are expected to be non--negligible and 
are sensitive to possible SUSY contributions~\cite{Mohanta-ACP-SUSY}.

\begin{table*}
\caption{\label{tab:summary} Results from the loose (top) and
 tight (bottom) selections.  
 Absolute branching fractions are normalized to the the world--average values
${\mathcal B}(\mbox{\BdKpi}) = (19.7\pm 0.6) \times 10^{-6}$ and
$f_{s}= (10.4 \pm 1.4)\%$ and $ f_{d}= (39.8 \pm 1.0)\%$~\cite{HFAG06}.
The first quoted uncertainty is statistical, the second is 
systematic. Limits are at 90\% CL.} 

{\footnotesize
\begin{tabular}{lc|lc|c}
\hline
mode & N$_{s}$ & Quantity & Measurement & \BR (10$^{-6}$)  \\
\hline
\BdKpi         & 4045 $\pm$ 84          & \acp               & -0.086 $\pm$ 0.023 $\pm$ 0.009   &                            \\ 
\Bdpipi        & 1121 $\pm$ 63          & \BdpipisuBdKpidef\ & 0.259 $\pm$ 0.017 $\pm$ 0.016    & 5.10 $\pm$ 0.33 $\pm$ 0.36 \\
\BsKK          & 1307 $\pm$ 64          & \BsKKsuBdKpidef\   &  0.324 $\pm$ 0.019 $\pm$ 0.041   & 24.4 $\pm$ 1.4 $\pm$ 4.6   \\
\hline
\BsKpi              & 230 $\pm$ 34 $\pm$ 16  & \BsKpisuBdKpidef\ &  0.066 $\pm$ 0.010 $\pm$ 0.010 & 5.0 $\pm$ 0.75 $\pm$ 1.0   \\
                    &                        &     \acp     &  0.39 		    $\pm$ 0.15 $\pm$ 0.08    &                 \\ 
\Bspipi             & 26 $\pm$ 16 $\pm$ 14   &\BspipisuBdKpidef\ &  
0.007 $\pm$ 0.004 $\pm$ 0.005 & 0.53 $\pm$ 0.31 $\pm$ 0.40 ($<1.36$)\\
\BdKK               & 61 $\pm$ 25  $\pm$ 35  & \BdKKsuBdKpidef\  &  
0.020 $\pm$ 0.008 $\pm$ 0.006 & 0.39 $\pm$ 0.16 $\pm$ 0.12 ($<0.7$)~~\\
 \LbpK               & 156 $\pm$ 20 $\pm$ 11  &\LbppisuLbpKdef\   &  0.66 $\pm$ 0.14 $\pm$ 0.08    &                            \\
 \Lbppi              & 110 $\pm$ 18 $\pm$ 16  &                   &                                &                            \\
\hline
\end{tabular}
}
\end{table*}

\subsection{CP asymmetries}

We can measure from our data the \CP\ asymmetries of both \Bd\ and 
\Bs\ decays in the self-tagging final state $K^{\pm}\pi^{\mp}$.
The asymmetry of the \Bs\ mode is measured with the tight
selection used in previous section, while the looser selection 
is used for the \Bd\ mode. 

The raw asymmetries returned by the fitting procedure need to be 
corrected for possible detector and procedural biases. 
This has been done by measuring the asymmetry in a sample of 1M 
prompt \DKpi\ decays, reconstructed and selected with the same computer code used for
analysis of the \Bhhgen\ sample and similar selection criteria. 
Given the smallness of the \CP\ asymmetry expected in the \DKpi\ mode
($<< 1\%$), the experimentally measured asymmetry
provides a good determination of the measurement bias, including 
asymmetries in the \dedx\ response or other possible unanticipated 
effects. The observed effect $(0.6 \pm 0.14)\%$ is in good 
agreement with the expected $K^{+}$/$K^{-}$ asymmetry due to the 
different probability of interaction with the detector material. 

The result $\acpbdkpi =-0.086 \pm 0.023 \pm 0.009$ is 
in agreement with the world--average~\cite{HFAG06}, and is the second 
most precise measurement.
The updated world average 
$A^{ave.}_{\mathsf{CP}}(\bdkpi)=-0.095 \pm 0.013$ has a significance 
of 7$\sigma$ (previously 6$\sigma$).
Comparison of this average with the asymmetry in the 
similar mode of the \Bu\ , shows a deviation of $4.8\sigma$ 
(previously $4.6\sigma$). While it has been argued in the past that 
the asymmetries in these two modes should be equal in the standard 
model due to isospin symmetry, this is not anymore considered a reliable 
test~\cite{Pierini,Mishima,Fleischer-CKM}. Conversely, it has 
been argued~\cite{Gronau:2000md,Lipkin-BsKpi} that a 
much more robust test of the Standard Model origin of the 
asymmetry of the \BdKpi\ mode is the comparison with the 
corresponding asymmetry in the \BsKpi\ mode, where the final state is 
identical, thus canceling possible effects of final state interactions. 
The predicted equality 
of rate differences $\Gamma(\aBdKpi)-\Gamma(\BdKpi) = 
\Gamma(\BsKpi)-\Gamma(\aBsKpi)$ is very robust under the Standard 
Model, while it would be completely fortuitous under a New Physics scenario, 
because it is produced by interference of very different amplitudes 
in the two cases.
In addition, the smallness of the \BR(\BsKpi) 
(Tab.~\ref{tab:summary}), makes the predicted 
asymmetry large ($\simeq 40\%$) and therefore experimentally more accessible.
Using our tight set of cuts, we find $\acpbskpi=0.39 \pm 0.15 \pm 0.08$.
This value favors the large \CP\ asymmetry predicted by the Standard 
Model and has the correct sign, but is still compatible with zero 
(significance just above $2\sigma$). By combining our measurement 
with the world-average for the \Bd\ we obtain
$\frac{\Gamma(\aBdKpi)-\Gamma(\BdKpi)}{\Gamma(\BsKpi)-\Gamma(\aBsKpi)}  
= 0.84 \pm 0.42 \pm 0.15$, in agreement with the Standard Model expectation 
of 1.0.
It will be very interesting to see if this agreement persists with more 
data. Given the large expected asymmetry, the SM asymmetry will be 
visible as a $3\sigma$ effect with 1.5~\lumifb\ of data, and a $5\sigma$ effect 
before reaching 4~\lumifb . A non observation of this asymmetry would 
indicate a non-SM source of \CP\ violation.

\subsection{Precision Branching Fractions}

The sample selection for \acpbdkpi\ also provides good measurements of the 
branching fractions of the `large yield' modes \BsKK\ and \Bdpipi . 
We obtain $\BR(\BsKK)= (24.4 \pm 1.4 \pm 4.6) \times 10^{-6}$, in 
agreement with current 
predictions~\cite{matiasBsKK,matiasBsKK2,B-N,Fleischer-CKM}
and with the previous CDF measurement \cite{paper_bhh}, although the 
lower central value now indicates a smaller U-spin breaking effect. 
Work is in progress to reduce the systematic uncertainty which dominates 
the resolution of the present preliminary results.

A substantial yield is also available in the \Bdpipi\ mode, allowing 
a measurement of the branching fraction: \BR(\Bdpipi)= ($5.10 \pm 0.33 
\pm 0.36)\times 10^{-6}$. This has the same precision, and is in agreement 
with the current results from \yqs ~\cite{HFAG06}.

\section{Prospects}

CDF results based on 1~\lumifb\ are beginning to show the Tevatron 
potential for B physics. The sample on tape is now almost doubled, 
and the luminosity keeps increasing; the default plan is to collect 8~\lumifb\ by year 2009. 
Most of the systematic uncertainty in the measurements described 
above is determined by statistical uncertainties in the calibration 
samples, and is therefore expected that the precision will keep 
increasing with statistics. Highlights from the full runII 
samples will be: \acp\ in \BdKpi\ at 1\% level; 5-sigma observation 
of direct \acp\ in \BsKpi\ 
(or alternatively the discovery of non-SM \CP\ violation) ; first \acp\  
measurements in the \Lb\ ; and tight constraint, or even observation, of annihilation modes.

In addition to the above, time-dependent measurements will be 
performed. Resolutions on time-dependent \CP\ asymmetries can be 
predicted from the yields, flavor tagger perfomance ($\epsilon 
D^{2}$), and effective S/B. Tagger performance was optimized to 
$\epsilon D^{2}=5.3\%$ for the \Bs\ mixing analysis; we assume a performance 
$\epsilon D^{2}=4\%$ can be obtained for the \Bd\ with similar 
methods. For the other parameters we assume the values of current 
analysis with no further improvements.

The \CP\ asymmetries in the \Bdpipi\ mode can be measured with 
a resolution $\sigma_{ACP}=0.15$ with 6~\lumifb , which will offer an 
interesting additional measurement of similar precision to the 
currently available results from \yqs\ , that still show a disagreement.

The performance in the measurement of \CP\ asymmetry in the \BsKK\ 
mode depends on the assumed effective S/B ($\simeq 1$, depending on the 
selection), and proper-time resolution (between 70 and 100 
fs$^{-1}$). A resolution in the range [0.15, 0.3] is expected for both direct and 
mixing \acp\ from a sample of 6~\lumifb . These asymmetries are related in the SM 
to the \Bd\ asymmetries and angle $\gamma$ via 
U-spin~\cite{Fleischer,Fleischer-CKM}, 
and are sensitive to possible SUSY effects~\cite{matiasBsKK2}.
In addition, it is possible to measure the \Bs\ lifetime in a \CP\ 
eigenstate, see~\cite{BsKK-life} for details.


\end{document}

%% file: macro.tex

\def\CPv     {$C\!P$}
\def\pep  {PEP-II}
\def\Hbabar{{\Huge\bf B}\hspace{-0.1em}{\LARGE\bf A}\hspace{-0.1em}{\Huge\bf B}\hspace{-0.1em}{\LARGE\bf A\hspace{-0.1em}R}\,}
\def\Lbabar{{\LARGE\sl B}\hspace{-0.1em}{\large\sl A}\hspace{-0.1em}{\LARGE\sl B}\hspace{-0.1em}{\large\sl A\hspace{-0.1em}R}\,}
\def\lbabar{{\sl {\Large\sl B}\hspace{-0.45em} A\hspace{-0.1em}{\Large\sl B}\hspace{-0.45em} A\hspace{-0.1em}R\,\,}}
\def\babarnos{\mbox{{\sl B\hspace{-0.4em} {\scriptsize\sl A}\hspace{-0.4em} B\hspace{-0.4em} {\scriptsize\sl A\hspace{-0.1em}R}}}}
\def\babarnote{\babar\ Note \#}
\newcommand{\comment}[1]{}
\def\LOI{{\sl Letter of Intent}}
\def\CDR{{\sl Conceptual Design Report}}

\def\BaBar{\babar}
%
%
%
%
%
%
\def \rightdownarrow
 {\kern.3em
 \rule[.5ex]{.15mm}{2ex}
 {\mbox{$\kern-0.1em{\longrightarrow}$}}
 }
\def\lessim{\mathrel {\vcenter {\baselineskip 0pt \kern 0pt
\hbox{$<$} \kern 0pt \hbox{$\sim$} }}}
\def\gessim{\mathrel {\vcenter {\baselineskip 0pt \kern 0pt
\hbox{$>$} \kern 0pt \hbox{$\sim$} }}}
%
%

%

\newcommand{\ket}[1]{\left| #1 \right\rangle}
\newcommand{\bra}[1]{\left\langle #1 \right|}
\newcommand{\CPp}{\textsf{CP}-parit\`a\ }

\newcommand{\beq}{\begin{equation}}
\newcommand{\eeq}{\end{equation}}
\newcommand{\bear}{\begin{array}}
\newcommand{\ear}{\end{array}}
\newcommand{\bet}{\begin{tabular}}
\newcommand{\eet}{\end{tabular}}
\newcommand{\beqn}{\begin{eqnarray}}
\newcommand{\eeqn}{\end{eqnarray}}

\newcommand{\bfh}{\begin{figure}[h]}
\newcommand{\efh}{\end{figure}[h]}
\newcommand{\cosi}{\mbox{cos\`{\i\ }}}
\newcommand{\Cosi}{\mbox{Cos\`{\i\ }}}
\newcommand{\dgr}{\mbox{$^0$}}
\newcommand{\mrad}{\mbox{mr}}
\newcommand{\nb}{\mbox{$\,nb$}}
\newcommand{\ubond}{$\mu$bond}
\newcommand{\fui}{\mbox{$\varphi$}}
\newcommand{\x}{\mbox{$x$}}
\newcommand{\y}{\mbox{$y$}}
\newcommand{\hth}{head-to-head}
\newcommand{\Hth}{Head-to-head}
\newcommand{\mum}{\mbox{$\mu$m}}
\newcommand{\bi}{\mbox{$b \ $}}
\newcommand{\ab}{\mbox{$\bar{b} \ $}}
\newcommand{\bb}{\mbox{$b\bar{b} \ $}}
\newcommand{\bad}{\mbox{$b\bar{d} \ $}}
\newcommand{\bau}{\mbox{$b\bar{u} \ $}}
\newcommand{\ba}{\mbox{$B$}}
\newcommand{\bab}{\mbox{$B\bar{B}$}}
\newcommand{\bz}{\mbox{$B^{0}$}}
\newcommand{\abz}{\mbox{$\bar{B^{0}}$}}
\newcommand{\bp}{\mbox{$B^{+}$}}
\newcommand{\besse}{\mbox{$B_{s}$}}
\newcommand{\ee}{\mbox{$e^{+}e^{-}$}}
\newcommand{\pp}{\mbox{$p\bar{p}\ $}}
\newcommand{\qq}{\mbox{$q\bar{q}\ $}}
\newcommand{\yqs}{\mbox{$\Upsilon(4S)$}}
\newcommand{\yts}{\mbox{$\Upsilon(3S)$}}
\newcommand{\yds}{\mbox{$\Upsilon(2S)$}}
\newcommand{\yus}{\mbox{$\Upsilon(1S)$}}
\newcommand{\fb}{\mbox{$f_B$}}
\newcommand{\mb}{\mbox{$m_B$}}
\newcommand{\fm}{\mbox{$f_M$}}
\newcommand{\ml}{\mbox{$m_l$}}
\newcommand{\ppiu}{\mbox{$p^+$}}
\newcommand{\npiu}{\mbox{$n^+$}}
\newcommand{\ubtn}{\mbox{$\bu \to \tau \nu_{\tau}$}}
\newcommand{\btn}{\mbox{$\ba \to \tau \nu$}}
\newcommand{\mbtn}{\mbox{$\bm \to \tau^- \bar{\nu}_{\tau}$}}
\newcommand{\pbtn}{\mbox{$\bp \to \tau^+ \nu_{\tau}$}}
\newcommand{\ts}{\mbox{$\tau$}}
\newcommand{\tp}{\mbox{$\tau ^+$}}
\newcommand{\tm}{\mbox{$\tau ^-$}}
\newcommand{\anu}{\mbox{$\bar{\nu}$}}
\newcommand{\atnu}{\mbox{$\bar{\nu} _{\tau}$}}
\newcommand{\tnu}{\mbox{$\nu _{\tau}$}}

\newcommand{\pder}[2]{frac{\partial #1}{\partial #2}}
\newcommand{\der}[2]{frac{d #1}{d #2}}

\newcommand{\eq}[1]{\ref{eq:#1}}




\newcommand{\tab}[1]{Tab.~\ref{tab:#1}}
\newcommand{\twotabs}[2]{Tabs.~\ref{tab:#1} and \ref{tab:#2}}
\newcommand{\tabs}[2]{Tabs.~\ref{tab:#1}--\ref{tab:#2}}
\newcommand{\threetabs}[3]{Tabs.~\ref{tab:#1}--\ref{tab:#2}--\ref{tab:#3}}

\newcommand{\cita}[1]{\cite{#1}}

\newcommand{\fig}[1]{Fig.~\ref{fig:#1}}
\newcommand{\twofig}[2]{Figs.~\ref{fig:#1} and \ref{fig:#2}}
\newcommand{\figs}[2]{Figs.~\ref{fig:#1}--\ref{fig:#2}}

\newcommand{\tev}{\ensuremath{\mathrm{Te\kern -0.1em V}}}
\newcommand{\gev}{\ensuremath{\mathrm{Ge\kern -0.1em V}}}	
\newcommand{\mev}{\ensuremath{\mathrm{Me\kern -0.1em V}}}	
\newcommand{\kev}{\ensuremath{\mathrm{ke\kern -0.1em V}}}	
\newcommand{\massgev}{\mbox{\gev/$c^2$}}			
\newcommand{\massmev}{\mbox{\mev/$c^2$}}			
\newcommand{\pgev}{\mbox{\gev/$c$}}				
\newcommand{\pmev}{\mbox{\mev/$c$}}				

\newcommand{\stat}{\ensuremath{\mathit{~(stat.)}}}		
\newcommand{\syst}{\ensuremath{\mathit{~(syst.)}}}		

\newcommand{\sye}[1]{\ensuremath{\pm #1}}	   		
\newcommand{\ase}[2]{\ensuremath{^{+ #1}_{- #2}}}		


\newcommand{\CPV}{\ensuremath{\mathsf{C\!P}}}			
\newcommand{\CP}{\ensuremath{\mathsf{CP}}}			
\newcommand{\C}{\ensuremath{\mathsf{C}}}			
					
\newcommand{\vij}{\ensuremath{V_{ij}}}				
\newcommand{\vud}{\ensuremath{V_{ud}}}
\newcommand{\vus}{\ensuremath{V_{us}}}
\newcommand{\vub}{\ensuremath{V_{ub}}}
\newcommand{\vcd}{\ensuremath{V_{cd}}}
\newcommand{\vcs}{\ensuremath{V_{cs}}}
\newcommand{\vcb}{\ensuremath{V_{cb}}}
\newcommand{\vtd}{\ensuremath{V_{td}}}
\newcommand{\vts}{\ensuremath{V_{ts}}}
\newcommand{\vtb}{\ensuremath{V_{tb}}}


\newcommand{\pap}{\proton\antiproton}		        	
\newcommand{\pt}{\ensuremath{p_{\rm{T}}}}			
\newcommand{\ptb}{\ensuremath{\pt(B)}}				
\newcommand{\ptbd}{\ensuremath{\pt(\bd)}}			
\newcommand{\ptbs}{\ensuremath{\pt(\bs)}}			


\newcommand{\geant}{{\scshape geant}}				
\newcommand{\bgen}{{\scshape bgenerator}}			
\newcommand{\pythia}{{\scshape pythia}}				

\newcommand{\minuit}{{\scshape minuit}}				
\newcommand{\migrad}{{\scshape migrad}}				
\newcommand{\hesse}{{\scshape hesse}}				


\newcommand{\proton}{\ensuremath{\rm{p}}}
\newcommand{\antiproton}{\ensuremath{\bar{\rm{p}}}}
\newcommand{\electron}{\ensuremath{\rm{e}^-}}
\newcommand{\positron}{\ensuremath{\rm{e}^+}}



\newcommand{\Y}{\ensuremath{\Upsilon}}
\newcommand{\Yunos}{\mbox{$\Upsilon$(1S)}}
\newcommand{\Ydues}{\mbox{$\Upsilon$(2S)}}
\newcommand{\Ytres}{\mbox{$\Upsilon$(3S)}}
\newcommand{\Yquattros}{\mbox{$\Upsilon$(4S)}}
\newcommand{\Ycinques}{\mbox{$\Upsilon$(5S)}}

\newcommand{\bd}{\ensuremath{B^{0}}}				
\newcommand{\bs}{\ensuremath{B^{0}_s}}				
\newcommand{\bu}{\ensuremath{B^{+}}}				
\newcommand{\bc}{\ensuremath{B^{+}_c}}				

\newcommand{\abd}{\ensuremath{\overline{B}^{0}}}		
\newcommand{\abs}{\ensuremath{\overline{B}^{0}_s}}		
\newcommand{\abu}{\ensuremath{B^{-}}}				
\newcommand{\abc}{\ensuremath{B^{-}_c}}				
	
\newcommand{\bphysics}{\mbox{$b$-physics}}

\newcommand{\bflavor}{\mbox{$b$-flavor}}			
\newcommand{\blavors}{\mbox{$b$-flavors}}			
										
\newcommand{\bquark}{\mbox{$b$-quark}}				
\newcommand{\bquarks}{\mbox{$b$-quarks}}			

\newcommand{\bhadron}{\mbox{$b$-hadron}}			
\newcommand{\bhadrons}{\mbox{$b$-hadrons}}			

\newcommand{\bbaryon}{\mbox{$b$-baryon}}			
\newcommand{\bbaryons}{\mbox{$b$-baryons}}			

\newcommand{\bg}{\ensuremath{B}}				
\newcommand{\bgmeson}{\mbox{$b$-meson}}				
\newcommand{\bgmesons}{\mbox{$b$-mesons}}			

\newcommand{\bn}{\ensuremath{B^{0}_{(s)}}}			
\newcommand{\bnmeson}{\mbox{$B^0_{(s)}$ meson}}			
\newcommand{\bnmesons}{\mbox{$B^0_{(s)}$ mesons}}		


\newcommand{\bhh}{\ensuremath{\bn \to h^{+}h^{'-}}}
\newcommand{\bdhh}{\ensuremath{\bd \to h^{+}h^{'-}}}
\newcommand{\bshh}{\ensuremath{\bs \to h^{+}h^{'-}}}
\newcommand{\bdpipi}{\ensuremath{\bd \to \pi^+ \pi^-}}
\newcommand{\bdkpi}{\ensuremath{\bd \to K^+ \pi^-}}
\newcommand{\abdkpi}{\ensuremath{\abd \to K^- \pi^+}}
\newcommand{\bskpi}{\ensuremath{\bs \to K^- \pi^+}}
\newcommand{\abskpi}{\ensuremath{\abs\to K^+ \pi^-}}
\newcommand{\bskk}{\ensuremath{\bs \to  K^+ K^-}}
\newcommand{\bspipi}{\ensuremath{\bs \to  \pi^+ \pi^-}}
\newcommand{\bdkk}{\ensuremath{\bd \to  K^+ K^-}}

\newcommand{\etal}{{\em et al.}}
\newcommand{\Vub}{$V_{ub}$}
\newcommand{\ppbar}{$\bar{p}p$}
\newcommand{\Bd}{$B^{0}$}
\newcommand{\BBd}{\ensuremath{\overline{B}^{0}}}
\newcommand{\Bu}{$B^{+}$}
\newcommand{\Bs}{$B_{s}^{0}$}
\newcommand{\Lb}{$\Lambda_{b}^{0}$}

\newcommand{\Bhh}{\ensuremath{\bn \to h^{+}h^{'-}}}
\newcommand{\Lbph}{\ensuremath{\Lambda^{0}_{b} \to ph^{-}}}
\newcommand{\Bdpipi}{\ensuremath{\bd \to \pi^+ \pi^-}}
\newcommand{\aBdpipi}{\ensuremath{\overline{B}^{0} \to \pi^+ \pi^-}}
\newcommand{\BdKpi}{\ensuremath{\bd \to K^+ \pi^-}}
\newcommand{\aBdKpi}{\ensuremath{\abd \to K^- \pi^+}}
\newcommand{\BsKpi}{\ensuremath{\bs \to K^- \pi^+}}
\newcommand{\aBsKpi}{\ensuremath{\abs\to K^+ \pi^-}}
\newcommand{\BsKK}{\ensuremath{\bs \to  K^+ K^-}}
\newcommand{\aBsKK}{\ensuremath{\overline{B}^{0}_{s} \to  K^+ K^-}}
\newcommand{\Bspipi}{\ensuremath{\bs \to  \pi^+ \pi^-}}
\newcommand{\aBspipi}{\ensuremath{\overline{B}^{0}_{s} \to  \pi^+ \pi^-}}
\newcommand{\BdKK}{\ensuremath{\bd \to  K^+ K^-}}
\newcommand{\aBdKK}{\ensuremath{\overline{B}^0 \to  K^+ K^-}}
\newcommand{\Lbppi}{\ensuremath{\Lambda_{b}^{0} \to p\pi^{-}}}
\newcommand{\aLbppi}{\ensuremath{\overline{\Lambda}_{b}^{0} \to \overline{p}\pi^{+}}}
\newcommand{\LbpK}{\ensuremath{\Lambda_{b}^{0} \to pK^{-}}}
\newcommand{\aLbpK}{\ensuremath{\overline{\Lambda}_{b}^{0} \to \overline{p}K^{+}}}

\newcommand{\bhha}{\ensuremath{\bn \to hh}}
\newcommand{\bdpipia}{\ensuremath{\bd \to \pi\pi}}
\newcommand{\bdkpia}{\ensuremath{\bd \to K\pi}}
\newcommand{\abdkpia}{\ensuremath{\abd \to K\pi}}
\newcommand{\bskpia}{\ensuremath{\bs \to K \pi}}
\newcommand{\abskpia}{\ensuremath{\abs\to K \pi}}
\newcommand{\bskka}{\ensuremath{\bs \to  K K}}
\newcommand{\bspipia}{\ensuremath{\bs \to  \pi\pi}}
\newcommand{\bdkka}{\ensuremath{\bd \to  KK}}
\newcommand{\Lbppia}{\ensuremath{\Lambda_{b}^{0} \to p\pi}}
\newcommand{\aLbppia}{\ensuremath{\overline{\Lambda}_{b}^{0} \to \overline{p}\pi}}
\newcommand{\LbpKa}{\ensuremath{\Lambda_{b}^{0} \to pK}}
\newcommand{\aLbpKa}{\ensuremath{\overline{\Lambda}_{b}^{0} \to \overline{p}}}

\newcommand{\Dhh}{\ensuremath{D^{0} \to h^{+}h^{'-}}}
\newcommand{\Dpipi}{\ensuremath{D^{0} \to \pi^+ \pi^-}}
\newcommand{\Dkpi}{\ensuremath{D^{0} \to K^- \pi^+}}
\newcommand{\aDkpi}{\ensuremath{\overline{D}^{0} \to K^+ \pi^-}}
\newcommand{\Dkk}{\ensuremath{D^{0} \to K^+ K^-}}

\newcommand{\DKpi}{\ensuremath{D^{0} \to K^- \pi^+}}
\newcommand{\aDKpi}{\ensuremath{\overline{D}^{0} \to K^+ \pi^-}}
\newcommand{\DKK}{\ensuremath{D^{0} \to K^+ K^-}}


\newcommand{\fracdedx}{\ensuremath{\frac{\rm{dE}}{\rm{dx}}}}
\newcommand{\dedx}{\ensuremath{dE/dx}}
\newcommand{\like}{\ensuremath{\mathscr{L}}}
\newcommand{\gauss}{\ensuremath{\mathscr{G}}}
\newcommand{\pdf}{\ensuremath{\wp}}
\newcommand{\ptot}{\ensuremath{p_{\rm{tot}}}}
\newcommand{\res}{\ensuremath{\delta}}
\newcommand{\id}{\ensuremath{\kappa}}

\newcommand{\acpbdkpi}{\ensuremath{A_{\mathsf{CP}}(\bdkpi)}}
\newcommand{\acpbskpi}{\ensuremath{A_{\mathsf{CP}}(\bskpi)}}
\newcommand{\acpLbppi}{\ensuremath{A_{\mathsf{CP}}(\Lbppi)}}
\newcommand{\acpLbpK}{\ensuremath{A_{\mathsf{CP}}(\LbpK)}}

\newcommand{\acpDKpi}{\ensuremath{A_{\mathsf{CP}}(\DKpi)}}
		      
\newcommand{\pipisukpi}{\ensuremath{\mathcal{B}(\bdpipi)/\mathcal{B}(\bdkpi)}}
\newcommand{\kksukpi}{\ensuremath{(\mathit{f_s}/\mathit{f_d})\cdot\mathcal{B}(\bskk)/\mathcal{B}(\bdkpi)}}
\newcommand{\pipisukk}{\ensuremath{(\mathit{f_d}/\mathit{f_s})\cdot\mathcal{B}(\bdpipi)/\mathcal{B}(\bskk)}}
\newcommand{\bskpisubdkpi}{\ensuremath{(\mathit{f_s}/\mathit{f_d})\cdot\mathcal{B}(\bskpi)/\mathcal{B}(\bdkpi)}}
\newcommand{\rateratio}{\ensuremath{(\mathit{f_d}/\mathit{f_s})\cdot 
(|A(\abdkpi)|^2 - |A(\bdkpi)|^2)/(|A(\abskpi)|^2 - |A(\bskpi)|^2)}}

\newcommand{\bspipisubdkpi}{\ensuremath{(\mathit{f_s}/\mathit{f_d})\cdot \mathcal{B}(\bspipi)/\mathcal{B}(\bdkpi)}}
\newcommand{\bdkksubdkpi}{\ensuremath{\mathcal{B}(\bdkk)/\mathcal{B}(\bdkpi)}}

\newcommand{\LbppisuBdKpi}{\ensuremath{\mathcal{B}(\Lbppi)/\mathcal{B}(\bdkpi)}}
\newcommand{\LbpKsuBdKpi}{\ensuremath{\mathcal{B}(\LbpK)/\mathcal{B}(\bdkpi)}}
\newcommand{\LbpKsuLbppi}{\ensuremath{\mathcal{B}(\LbpK)/\mathcal{B}(\Lbppi)}}
\newcommand{\LbppisuLbpK}{\ensuremath{\mathcal{B}(\Lbppi)/\mathcal{B}(\LbpK)}}

\newcommand{\Nbskpi}{\ensuremath{N(\bskpi)}}
\newcommand{\Nbspipi}{\ensuremath{N(\bspipi)}}
\newcommand{\Nbdkk}{\ensuremath{N(\bdkk)}}
\newcommand{\NLbppi}{\ensuremath{N(\Lbppi)}}
\newcommand{\NLbpK}{\ensuremath{N(\LbpK)}}

\newcommand{\limbskpi}{\ensuremath{(\mathit{f_s}/\mathit{f_d})\cdot\mathcal{B}(\bskpi)/\mathcal{B}(\bdkpi)}}
\newcommand{\limbspipi}{\ensuremath{\mathcal{B}(\bspipi)/\mathcal{B}(\bskk)}}
\newcommand{\limbdkk}{\ensuremath{\mathcal{B}(\bdkk)/\mathcal{B}(\bdkpi)}}

\newcommand{\fracpipisukpi}{\ensuremath{\frac{\mathcal{B}(\bdpipi)}{\mathcal{B}(\bdkpi)}}}
\newcommand{\frackksukpi}{\ensuremath{\frac{\mathit{f_s}}{\mathit{f_d}}\cdot\frac{\mathcal{B}(\bskk)}{\mathcal{B}(\bdkpi)}}}
\newcommand{\fracpipisukk}{\ensuremath{\frac{\mathit{f_d}}{\mathit{f_s}}\cdot\frac{\mathcal{B}(\bdpipi)}{\mathcal{B}(\bskk)}}}
\newcommand{\fraclimbskpi}{\ensuremath{\frac{\mathit{f_s}}{\mathit{f_d}}\cdot\frac{\mathcal{B}(\bskpi)}{\mathcal{B}(\bdkpi)}}}
\newcommand{\fraclimbspipi}{\ensuremath{\frac{\mathcal{B}(\bspipi)}{\mathcal{B}(\bskk)}}}
\newcommand{\fraclimbdkk}{\ensuremath{\frac{\mathcal{B}(\bdkk)}{\mathcal{B}(\bdkpi)}}}

\newcommand{\BRbdpipi}{\ensuremath{\mathcal{B}(\bdpipi)}}
\newcommand{\BRbskk}{\ensuremath{\mathcal{B}(\bskk)}}
\newcommand{\BRbspipi}{\ensuremath{\mathcal{B}(\bspipi)}}
\newcommand{\BRbskpi}{\ensuremath{\mathcal{B}(\bskpi)}}
\newcommand{\BRbdkk}{\ensuremath{\mathcal{B}(\bdkk)}}

\newcommand{\BR}{\ensuremath{\mathcal B}}
\newcommand{\cdf}{CDF Collaboration}
\newcommand{\babar}{BaBar Collaboration}
\newcommand{\cleo}{CLEO Collaboration}
\newcommand{\belle}{Belle Collaboration}
\newcommand{\alep}{ALEPH Collaboration}

\newcommand{\ACPddef}{\ensuremath{{\frac{\BR (\aBdKpi)-\BR 
(\BdKpi)}{\BR (\aBdKpi)+\BR (\BdKpi)}}}}
\newcommand{\ACPsdef}{\ensuremath{{\frac{\BR (\aBsKpi)-\BR 
(\BsKpi)}{\BR (\aBsKpi)+\BR (\BsKpi)}}}}
\newcommand{\rateratiodef}{\ensuremath{\frac{\mathit{f_d}}{\mathit{f_s}}{\frac{ \Gamma(\aBdKpi)- 
\Gamma(\BdKpi)}{\Gamma(\aBsKpi)-\Gamma(\BsKpi)}}}}

\newcommand{\BdpipisuBdKpidef}{\ensuremath{\frac{\BR(\Bdpipi)}{\BR(\BdKpi)}}}
\newcommand{\BsKKsuBdKpidef}{\ensuremath{\frac{\mathit{f_s}}{\mathit{f_d}}\frac{\BR(\BsKK)}{\BR(\BdKpi)}}}
\newcommand{\BdpipisuBsKKdef}{\ensuremath{\frac{\mathit{f_d}}{\mathit{f_s}}\frac{\BR(\Bspipi)}{\BR(\BsKK)}}}
\newcommand{\BsKpisuBdKpidef}{\ensuremath{\frac{\mathit{f_s}}{\mathit{f_d}}\frac{\BR(\BsKpi)}{\BR(\BdKpi)}}}
\newcommand{\BspipisuBdKpidef}{\ensuremath{\frac{\mathit{f_s}}{\mathit{f_d}}\frac{\BR(\Bspipi)}{\BR(\BdKpi)}}}
\newcommand{\BdKKsuBdKpidef}{\ensuremath{\frac{\BR(\BdKK)}{\BR(\BdKpi)}}}

\newcommand{\LbppisuBdKpidef}{\ensuremath{\frac{\mathit{f_{\Lambda}}}{\mathit{f_d}}\frac{\BR(\Lbppi)}{\BR(\BdKpi)}}}
\newcommand{\LbpKsuBdKpidef}{\ensuremath{\frac{\mathit{f_{\Lambda}}}{\mathit{f_d}}\frac{\BR(\LbpK)}{\BR(\BdKpi)}}}
\newcommand{\LbppisuLbpKdef}{\ensuremath{\frac{\BR(\Lbppi)}{\BR(\LbpK)}}}

\newcommand{\ie}{i.e.}					
\newcommand{\eg}{e.g.}					
 
\newcommand{\lumi}{\mbox{cm$^{-2}$s$^{-1}$}}  	 		
\newcommand{\Lumi}{\ensuremath{\mathcal{L}}}			
\newcommand{\lumipb}{\mbox{pb$^{-1}$}}				
\newcommand{\lumifb}{\mbox{fb$^{-1}$}}				

\newcommand{\br}{\ensuremath{\mathcal{B}}}
\newcommand{\acp}{\ensuremath{A_{\mathsf{CP}}}}

\newcommand{\no}{\nonumber}
\newcommand{\cO}{{\cal O}}
\newcommand{\cA}{{\cal A}}

%% file: ckm2006_punzi.bbl
\begin{thebibliography}{99}
\bibitem{C-conjugate} \C-conjugate modes
are implied and branching fractions intended as
\CP -averages, unless otherwise stated.
\bibitem{isolation} This is defined by $I_{B}= \ptb/[\ptb + \sum_{i} \pt(i)]$, 
where the sum runs over every other track within
a cone of radius one in the $\eta-\phi$ space around the $\Bhad$ flight direction.
\end{thebibliography}
